\documentclass[12pt]{article}

\usepackage{amsmath}
\usepackage{amssymb}
\usepackage{amsfonts}
\usepackage{graphicx}

\numberwithin{equation}{section}

\begin{document}

\begin{center}
{\large \bf{ The Higgs condensate as a quantum liquid: Comparison with the  full Run 2  CMS data }}
\end{center}

\vspace*{2cm}

\begin{center}
{
Paolo Cea~\protect\footnote{Electronic address:
{\tt paolo.cea@ba.infn.it}}  \\[0.5cm]
{\em INFN - Sezione di Bari, Via Amendola 173 - 70126 Bari,
Italy} }
\end{center}

\vspace*{1.5cm}

\begin{abstract}
\noindent 
We compare our  proposal for  an additional heavy Standard Model Higgs boson to the available  
full data set collected by the CMS detector during Run 2 at the Large Hadron Collider (LHC)  corresponding to an integrated luminosity 
of 138 fb$^{-1}$. The CMS Collaboration performed a search for a high mass Higgs boson decaying into a pair of W bosons 
in the dileptonic channel. Our analysis of the CMS data indicated  the presence of a broad excess in the mass range
600 GeV - 800 GeV  with respect  to the   expected Standard Model background with a rather significative statistical significance.
We found that  our theoretical proposal is in  reasonable agreement  with the  experimental  observations.
\vspace{0.5cm}

\vspace{0.2cm}


\end{abstract}
\newpage
\noindent
\section{Introduction}
\label{s-1}
In 2012 a new particle was discovered by the ATLAS and CMS experiments at the LHC~\cite{Aad:2012,Chatrchyan:2012}.
This new particle, with a mass of about 125 GeV, is consistent with the Standard Model 
Higgs boson. With this discovery the Higgs mechanism~\cite{Englert:1964,Higgs:1964,Guralnik:1964,Higgs:1966}
has been validated. 
The Standard Model of Particle Physics introduces the Higgs mechanism in order to explain particle masses by means of the
so-called spontaneous symmetry breaking mechanism. Spontaneous symmetry breaking can be obtained by introducing a scalar field, the Higgs field, with a specific potential. Actually, the Higgs mechanism had emerged as the only  mechanism capable of reconciling gauge field theories
 with the observed mass spectrum. Indeed, the striking conceptual and empirical success of the Standard Model established 
 increasing trust in the viability of the Higgs mechanism. \\
Usually the spontaneous symmetry breaking in the Standard Model is implemented
within the perturbation theory which leads to predict that the Higgs boson mass squared  is proportional to $\lambda \, v^2$,
where $\lambda$ is the renormalised scalar self-coupling and  $v \simeq 246$  GeV  is the known weak scale. 
 On the other hand, it is known that,  within the non-perturbative  description of spontaneous symmetry breaking,
 self-interacting scalar fields are subject to the triviality problem~\cite{Fernandez:1992}, namely the renormalised self-coupling
$\lambda \rightarrow  0$ when the ultraviolet cutoff  is sent to infinity.  Strictly speaking, there are  no rigorous proof 
of triviality~\footnote{It is worthwhile to mention interesting progresses  in this direction offered by the recent proof of the Gaussian structure
of the scaling limits of the  critical  Ising and $\phi^4$ fields in the marginal case of four  dimensions~\cite{Aizenman:2021}.}.
On the other hand, there exist several numerical studies  which leave little doubt on the triviality conjecture.  As a consequence, 
within the perturbative approach,  the scalar sector of the Standard Model  must be considered just an effective description 
 valid only up to some (unknown) cut-off scale. 
Notwithstanding, extensive numerical simulations showed that even without self-interactions the scalar bosons could trigger spontaneous
 symmetry breaking. Moreover, precise non-perturbative numerical simulations~\cite{Cea:2004,Cea:2012} indicated that the excitation
of the Bose-Einstein scalar condensate should be  a rather heavy scalar particle with mass of about 750(30) GeV~\cite{Cea:2012}. \\
To reconcile the overwhelming evidence of a rather light Higgs boson with mass of 125 GeV (indicated with h) with
the indications from non-perturbative studies of a heavy Higgs boson with mass around 750 GeV (indicated with H),
in a recent article~\cite{Cea:2020} we advanced the proposal that  the Higgs condensate of the Standard Model  should be considered
like a  relativistic quantum fluid analogous  to superfluid helium. As discussed at length in Ref.~\cite{Cea:2020}, we found that
 there are two different kind of Higgs condensate excitations that are similar to phonon and rotons in He II. Moreover,
 in the dilute gas approximation these two Higgs condensate excitations behave like the  perturbative Standard Model  Higgs 
 boson and a heavy Standard Model  Higgs boson. 
Indeed, in Ref.~\cite{Cea:2019} we presented  some  phenomenological consequences of the Standard Model heavy Higgs boson
proposal.  In particular, we discussed  the couplings of the H Higgs boson to the massive vector bosons and to fermions,
 the expected production mechanisms, and the main decay modes.   We also attempted a quantitative comparison in the so-called 
 golden channel  $H \rightarrow \ell^+ \ell^- \ell'^+ \ell'^-$ where  $\ell , \ell' = e$ or $\mu$.
 More precisely,  by means of an unofficial combination of  the preliminary Run 2 data collected by the  ATLAS and CMS 
 experiments,  we found some evidence  of a broad scalar resonance  that looked  consistent 
with our Standard Model heavy Higgs boson~\cite{Cea:2019,Cea:2020}. Unfortunately, this preliminary
evidence has not been corroborated  by the full data sets collected by the ATLAS Collaboration during the LHC Run 2.
 In Ref.~\cite{Cea:2021} we  critically compared our proposal  to the full Run 2 data sets released  by the ATLAS Collaboration. 
 A search for a new high-mass resonance  has been performed by the ATLAS experiment using data collected at $\sqrt{s}$  = 13 TeV
 corresponding to an integrated luminosity of 139 fb$^{-1}$ both  in the golden channel  $H \rightarrow \ell^+ \ell^- \ell'^+ \ell'^-$,   
 $\ell , \ell' = e$ or $ \mu$, and in the decays into WW or ZZ with production mechanisms and  branching ratios that mimic the ones 
 of a heavy Standard Model Higgs boson. We do not found a clear statistical evidence for our  heavy Higgs boson. 
 At beast we found a hint of a signal in the gluon-gluon fusion Higgs production mechanism.
 As a matter of fact, we reached the conclusion that  there 
  was not enough sensitivity to detect the signal in vector-boson fusion mechanism mainly due to tight event-selection cuts.
 In any case, we   concluded  that our theoretical proposal was still not ruled out by  the ATLAS observations. 
However, we must admit that it is problematic the absence of experimental evidence in the  decays of the heavy Higgs boson into
two W vector bosons. It is known that the main decay mode of a heavy Higgs boson is the decay into WW. 
Therefore,  the most stringent constraints should arise from the experimental searches for a heavy Higgs boson decaying into two W gauge bosons.
Fortunately, the CMS Collaboration recently  reported a  search for high-mass resonances decaying into a pair of W bosons into
 the fully leptonic final state using the full Run 2 data set corresponding to an integrated luminosity of 
 138 fb$^{-1}$~\cite{CMS:2022}.  The aim of the  present note is to contrast our theoretical expectations  to the LHC Run 2 data  
 from the CMS Collaboration in the above specified  decay channel.  We will show that the CMS data display a broad excess
 that seems to compare favourably with our proposal. \\
This paper is organised as follows. In Sect.~\ref{s-2}, for the  reader's convenience, we briefly discuss our theoretical
proposal for two Standard Model Higgs bosons together with  the main production mechanisms and decay modes.
Sect.~\ref{s-3} is devoted to the comparison with the latest available CMS data.
Finally, in Sect.~\ref{s-4} we summarise the main result of the present paper and draw our conclusions.
\section{Heavy and light  Higgs bosons}
\label{s-2}
In the present Section we would like to illustrate very  briefly  the proposal  advanced in Ref.~\cite{Cea:2020} to look at the Higgs condensate 
as a quantum liquid analogous to the Bose-Einstein condensate in superfluid He II. We found  that the low-lying excitations of the Higgs condensate 
 resembled two Higgs bosons that  were the relativistic version of the phonons and rotons in superfluid He II. 
   Actually,  in the dilute gas approximation that is 
the relevant regime for the LHC physics, these  low-lying excitations of the Higgs condensate resembled  two  Standard Model Higgs bosons
 with masses around  100 GeV and  750 GeV, respectively.
The lighter Higgs boson h was identified with the new LHC particle with  mass $M_h \simeq 125$ GeV that seemed  to behave
 consistently with the Standard Model perturbative Higgs boson.  On the other hand, the heavy Standard Model Higgs boson H, in accordance with
 the phenomenological analyses presented in Ref.~\cite{Cea:2019}, was assumed to have mass $M_H \simeq 730$ GeV.
Note that this mass value is in  accordance  with previous extensive numerical studies~\cite{Cea:2004,Cea:2012}.
We emphasise  that we are not saying that there are two different elementary quantum Higgs fields.
On the contrary, we have a unique quantum Higgs field. However, since the scalar condensate
behaves like the He II quantum liquid, when the Higgs field acts on the condensate
it can give rise to two elementary excitations, namely the phonon-like and roton-like excitations
corresponding to long-range collective and localised disturbances of the condensate, respectively.
These elementary condensate excitations behave as weakly interacting scalar
fields with vastly different mass. Moreover,  one remarkable aspect of  our approach is that the Higgs boson
masses are not free parameters, but these can be estimated from first principles. \\
Once established that  the perturbations of the scalar condensate due to the quantum Higgs field behave as two independent
weakly interacting massive scalar fields, we need to investigate  the experimental signatures  and the interactions of these
Higgs  condensate elementary excitations.  Obviously, the most striking consequence of
our approach is the prevision  of an additional  heavy  Higgs boson.  As we already said, the light Higgs boson is the
natural candidate for the new LHC scalar resonance at 125 GeV. On the other hand, our previous phenomenological analysis of the
preliminary LHC Run 2 data in the golden channel~\cite{Cea:2019} corroborated  the presence of a broad scalar resonance with
central mass at 730 GeV.   These two Higgs bosons will interact with the gauge vector bosons.
We already pointed out~\cite{Cea:2020,Cea:2019} that the couplings of the Higgs condensate elementary excitations to the
gauge vector bosons are fixed by the gauge symmetries. As a consequence, both the Higgs bosons h and H  will be
coupled to gauge bosons as in the usual perturbative approximation of the Standard Model.  Moreover, these scalar
bosons have an effective finite self-coupling $\lambda_{eff}$ that, in general, is smaller than  the perturbative
 renormalised scalar self-coupling $\lambda$.
As concern the coupling to fermion fields, if we admit the presence of the Yukawa terms in the Lagrangian, then we are led to an effective
Yukawa lagrangian:
\begin{equation}
\label{2.1}
{\cal L}_{Y}^{eff}(x) = 
\sqrt{Z^h_{wf}}  \; \frac{\lambda_f}{\sqrt{2}} \;  \hat{\bar{\psi}}_f(x)  \,   \hat{\psi}_f(x) \;   \hat{h}(x)   +  
\sqrt{Z^H_{wf}} \;  \frac{\lambda_f}{\sqrt{2}}  \; \hat{\bar{\psi}}_f(x)  \,  \hat{\psi}_f(x) \;  \hat{H}(x)  \; , 
\end{equation}
where  $\hat{\psi}_f(x)$ indicates  a generic fermion quantum field and  the Yukawa coupling satisfies the usual relation:
\begin{equation}
\label{2.2}
\lambda_f \; = \;  \frac{\sqrt{2} \, m_f }{v} \; .
\end{equation}
In  Eq.~(\ref{2.1})  $Z^h_{wf}$ and   $Z^H_{wf}$ are wavefunction renormalisation constant that, roughly,  take care
of the eventual mismatch in the overlap between the fermion and quasiparticle wavefunctions. 
A direct calculation of the wavefunction renormalisation constants is not easy. Nevertheless,  in Ref.~\cite{Cea:2020} 
we fixed  these constants from a  comparison with the experimental observations.  As a  result we argued  
that~\footnote{The wavefunction renormalisation constant $Z^H_{wf}$ coincides with the phenomenological parameter $\kappa$
introduced in Ref.~\cite{Cea:2019}.}:
\begin{equation}
\label{2.3}
Z^h_{wf}  \; \simeq \; 1 \;  \; \; , \; \; \;  Z^H_{wf} \; \simeq  \; \frac{m_h}{m_H} \; .
\end{equation}
Note that   Eq.~(\ref{2.3}) has the remarkable consequence  that our light Higgs boson h is practically indistinguishable from the perturbative Higgs boson.  As a consequence, in the following we shall concentrate on the hypothetical heavy Higgs boson. \\
In our previous papers~\cite{Cea:2020,Cea:2019} we argued that for large Higgs masses the main
production processes are by vector-boson fusion and gluon-gluon fusion processes.
To evaluate the Higgs event production at LHC we need the inclusive Higgs production cross section. As in perturbation
theory, for large Higgs masses the main production processes are by vector-boson fusion (VBF)  and gluon-gluon fusion (GGF). 
In fact, since the couplings of the H boson to vector bosons are the same as those of a Standard Model Higgs boson, the H boson
 production cross section by vector-boson fusion is the same as in the perturbative Standard Model calculations.
As concern the gluon fusion production mechanism,  it is known that  the gluon coupling to the Higgs boson in the Standard Model is 
mediated mainly  by triangular loops of top and bottom quarks. Indeed,  in perturbation theory the Yukawa couplings 
of the Higgs particle to heavy quarks grows with quark mass, thus balancing the decrease of the triangle amplitude so that
the effective gluon coupling  approaches a non-zero value for large loop-quark masses. This means that for
heavy Higgs the gluon fusion inclusive cross section is almost completely  determined by the top quark.
Therefore, the total inclusive cross section for the production of the H Higgs boson
can be written as:
\begin{equation}
\label{2.4}
\sigma(p \; p \;  \rightarrow \; H) \; \simeq \;    \sigma_{VBF}(p \; p \;  \rightarrow \; H)
\; + \;  \sigma_{GGF}(p \; p \;  \rightarrow \; H)  \; ,
\end{equation}
where  $\sigma_{VBF}$ and   $\sigma_{GGF}$ are the vector-boson fusion and gluon-gluon fusion inclusive cross
sections, respectively. 
In the Standard Model the calculations of the cross sections computed at next-to-next-to-leading 
and next-to-leading order  for  a high mass Higgs boson with Standard Model-like couplings
 at $\sqrt{s} = $ 13  TeV are provided by the LHC Higgs Cross Section Working Group~\cite{Hcross13Tev}.
As concern the Standard Model gluon-gluon fusion cross section  we found~\cite{Cea:2019} that  this cross section can be
 usefully interpolated by:
\begin{equation}
\label{2.5}
 \sigma_{GGF}^{SM}(p \; p  \rightarrow H)  \; \simeq \;  
 \left\{ \begin{array}{ll}
 \;  \left (  \frac{ a_1}{ M_H} 
 \; + \; a_2 \; M_{H}^3  \right )  \;  \exp (-  a_3 M_{H})  \; \; &  M_{H}  \; \leq \; 300  \; GeV 
  \\
 \; \; \;  \; a_4  \;  & 300 \; GeV    \leq  M_{H}   \leq  400 \; GeV
  \\
 \; \;  \; \;a_4 \;  \exp \big [ - a_5 ( M_{H} - 400 \; GeV) \big ]  \; \; &  400  \; GeV \; \leq \; M_{H}
\end{array}
    \right.
\end{equation}
where $M_{H} $  is expressed in  GeV and
\begin{eqnarray}
\label{2.6}
a_1 \simeq 1.24 \, 10^4 \; pb \, GeV \;  \; , \; \;  a_2 \simeq 1.49 \, 10^{-6} \; pb \, GeV^{-3} \; , \;  
\nonumber \\
a_3 \simeq 7.06 \, 10^{-3} \;  GeV^{-1} \; , \;  \; a_4 \simeq 9.80 \;\, pb  \; , \hspace{2.2 cm}
\\ \nonumber
a_5 \simeq 7.63 \, 10^{-3}  \; GeV^{-1}  \; . \hspace{5.05 cm}
\end{eqnarray}
Likewise   the Standard Model vector-boson fusion cross section  can be parametrised as:
\begin{equation}
\label{2.7}
 \sigma_{VBF}^{SM}(p \; p  \rightarrow  H) \; \simeq \;    \bigg ( b_1 \; + \;  \frac{ b_2}{ M_{H}} 
 \; + \; \frac{b_3}{ M_{H}^2}  \bigg )  \;  \exp (-  b_4 \;  M_{H} )   \; ,
\end{equation}
with:
\begin{eqnarray}
\label{2.8}
b_1 \simeq - 2.69 \, 10^{-6}  \; pb   \;  \; , \; \;  b_2 \simeq 8.08 \, 10^{2} \; pb \, GeV \; , \hspace{1.15 cm}
 \nonumber \\
b_3 \simeq - 1.98 \, 10^{4}  \; pb \,  GeV^{2}  \; \;  , \;  b_4 \simeq  2.26 \, 10^{-3} \; GeV^{-1} \; .  \; \; \,
\end{eqnarray}
\begin{figure}
\vspace{-0.5cm}
\begin{center}
\includegraphics[width=0.7\textwidth,clip]{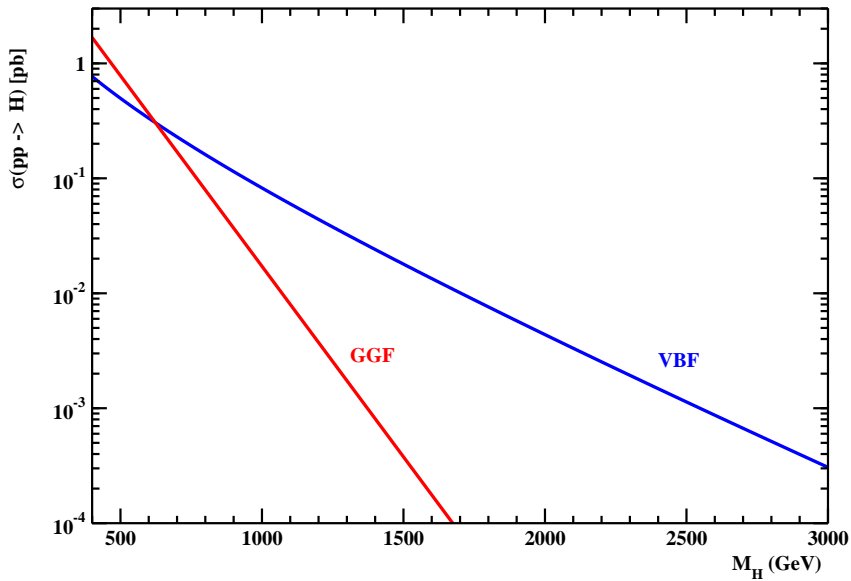}
\end{center}
\caption{\label{Fig1} 
Inclusive H Higgs boson  production cross sections  for the VBF  processes (blue continuous line) and
GGF processes (red continuous line) as a function of $M_H$ at $\sqrt{s} = $ 13  TeV.}
\end{figure}
Our previous discussion lead us to assume that to a  good approximation we can write:
\begin{equation}
\label{2.9}
  \sigma_{VBF}(p  p   \rightarrow  H)  \simeq   \sigma_{VBF}^{SM}(p p   \rightarrow  H) \;   \; , \; \; 
  \sigma_{GGF}(p p  \rightarrow  H)  \simeq   Z^H_{wf} \;  \sigma_{GGF}^{SM}(p p  \rightarrow  H) \; .
\end{equation}
In Fig.~\ref{Fig1} we compare the VBF  and GGF production cross sections given by Eq.~(\ref{2.9}) after taking into
account the value of $Z^H_{wf}$ in Eq.~(\ref{2.3}). From Fig.~\ref{Fig1} we see that the main production mechanism
of the heavy H Higgs boson is  by the VBF  processes since:
\begin{equation}
\label{2.10}
  \sigma_{VBF}(p  p   \rightarrow  H) \;  \simeq \;   2 \;   \sigma_{GGF}(p p  \rightarrow  H)   \;   \;  , \;  \;  M_H  \; \simeq \; 730 \;  GeV \; .
\end{equation}
In order to determine the phenomenological signatures of the massive H Higgs boson we need to examine the decay modes.
Given the rather large mass of the heavy Higgs boson, the main decay modes are the decays into two massive
vector bosons (see, e.g., Refs.~\cite{Gunion:1990,Djouadi:2008}):
\begin{equation}
\label{2.11}
\Gamma( H \; \rightarrow \; W^+ \, W^-)  \; \simeq  \;  \frac{G_F \, M^3_{H}}{8 \pi \sqrt{2}} \;
 \sqrt{1 - \frac{4 m^2_W}{M^2_{H}}} \;  \bigg ( 1 - 4 \,  \frac{m^2_W}{M^2_{H}} + 12 \, \frac{ m^4_W}{M^4_{H}}
 \bigg ) \; 
\end{equation}
and
\begin{equation}
\label{2.12}
\Gamma( H  \;  \rightarrow \;  Z^0 \, Z^0) \; \simeq \;   \frac{G_F \, M^3_{H}}{16 \pi \sqrt{2}} \;
 \sqrt{1 - \frac{4 m^2_Z}{M^2_{H}}} \; \bigg ( 1 - 4 \,  \frac{m^2_Z}{M^2_{H}} + 12 \, \frac{ m^4_Z}{M^4_{H}}
 \bigg )  \; . 
\end{equation}
The couplings of the H Higgs boson to the fermions are given by the Yukawa couplings. For heavy Higgs the only relevant fermion coupling is 
the top Yukawa coupling. The width for the decays of the H boson into a $t \bar{t}$ pairs is easily found~\cite{Gunion:1990,Djouadi:2008}:
\begin{equation}
\label{2.13}
\Gamma( H \rightarrow \; t \, \bar{t}) \; \simeq \; Z^H_{wf}  \;  \frac{3 \, G_F \, M_{H} m^2_t}{4 \pi \, \sqrt{2}} \;
\bigg ( 1 - 4 \,  \frac{m^2_t}{M^2_{H}}  \bigg )^{\frac{3}{2}}  \; ,
\end{equation}
where we have taken into account Eq.~(\ref{2.1}). So that, to a good approximation, the heavy Higgs  boson total width is given by:
\begin{equation}
\label{2.14}
\Gamma_{H}  \; \simeq \; \Gamma( H \rightarrow W^+ \, W^-)  \; + \; \Gamma( H \rightarrow Z^0 \, Z^0)  \; + \;
 \Gamma( H \rightarrow t \, \bar{t})  \; .
\end{equation}
\section{Comparison with the CMS Run 2 dataset}
\label{s-3}
The discussion in the previous Section showed that  our heavy Higgs boson is a rather broad resonance 
and that almost all the  decay modes  are given by the decays into W$^+$W$^-$ and Z$^0$Z$^0$ with:
\begin{equation}
\label{3.1}
Br(H \rightarrow W^+ W^-) \; \simeq \; 2 \; Br(H \rightarrow Z^0 Z^0) \; .
\end{equation}
In our previous papers  we found some evidence  of a broad scalar resonance  that looks  consistent 
with our Standard Model heavy Higgs boson in the golden channel.  Actually, the decay channels $H  \rightarrow  ZZ  \rightarrow  4 \ell$ 
 have very low branching ratios, but  the presence of leptons allows to efficiently reduce the huge background due mainly to diboson production.
In fact, the four-lepton channel, albeit rare, has the clearest and cleanest signature of all the
possible Higgs boson decay modes due to the small background contamination. 
Nevertheless,  from one hand we did not find  convincingly evidence of a heavy Standard Model Higgs boson in the
comparison with the full LHC Run 2 data sets released by the ATLAS Collaboration. On the other hand, according to
Eq.~(\ref{3.1}),  the main decay mode of a heavy Higgs boson is the decays into two
 W vector bosons. As a consequence, the most stringent constraints should arise from the experimental
 searches for a heavy Higgs boson decaying into two W gauge bosons. The lack,   at least up to now, of experimental evidences 
in this decay channel looks problematic. In fact, if this situation should persist our theoretical
proposal should not be  in agreement with observations and, therefore, should be rejected. 
The  aim of the present Section is to contrast our proposal with the recent document Ref.~\cite{CMS:2022} where
 the CMS Collaboration presented a search for a high mass resonance decaying into a pair 
of W bosons, using the full data set recorded by CMS during the LHC Run 2
corresponding to an integrated luminosity of 138 fb$^{-1}$.
 The search strategy for $H \rightarrow W^+W^-$ was based on the final state in which
  both   W bosons decay leptonically, resulting in a signature with two isolated, 
   oppositely   charged, high $p_T$ leptons (electrons or muons) and large missing 
  transverse  momentum,  due to the undetected neutrinos. So that,
  the bulk of the signal comes  from direct W decays to electrons or muons of opposite 
  charge. However, even if not explicitly mentioned in Ref.~\cite{CMS:2022}, the small contributions 
   proceeding through an intermediate $\tau$ lepton are implicitly included.
 Therefore,  in Ref.~\cite{CMS:2022} it is always included  the $W$ boson decays into all three lepton types,  
 so that the $\ell$ symbol in $W \rightarrow \ell + \nu$  comprises all three leptons
  $e, \mu, \tau$.
To increase the signal sensitivity event categorisation optimised for the gluon-gluon fusion  and
vector-boson fusion  production mechanisms were used. To this end,  it was introduced
a parameter $f_{VBF}$ corresponding to the fraction of the VBF production cross section 
  with respect to the total cross section. In this way,  $f_{VBF} = 0$ corresponds to GGF production signal,
  while  $f_{VBF} = 1$  considers only the VBF production signal. For a heavy scalar resonance
 with Standard Model-like couplings  $f_{VBF}$  was  set to the expectation using
 the cross sections provided by the LHC Higgs Cross SectionWorking Group~\cite{Hcross13Tev}.
\begin{figure}
\vspace{-0.5cm}
\begin{center}
\includegraphics[width=0.7\textwidth,clip]{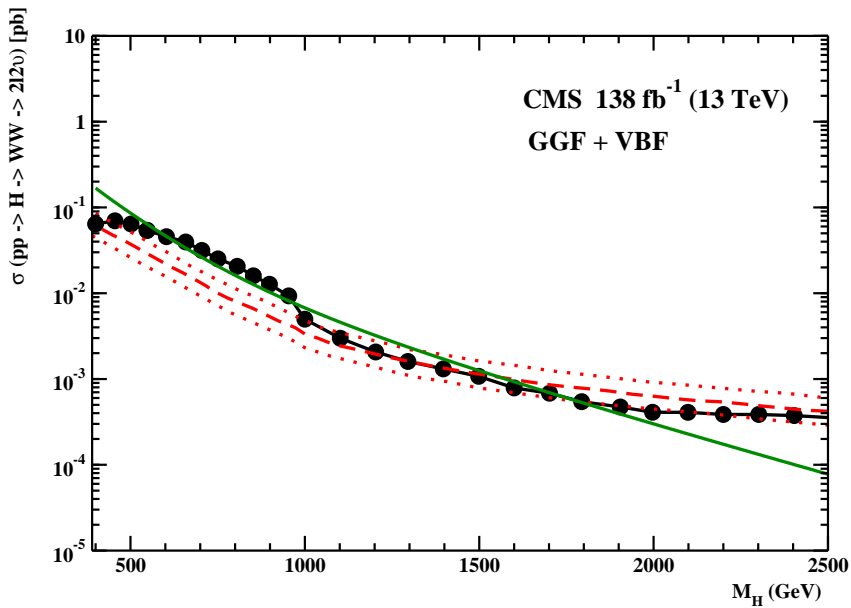}
\end{center}
\caption{\label{Fig2} 
Product of the cross section   $\sigma(p  p   \rightarrow  H)$  with  the  branching ratio  $Br(H \rightarrow WW \rightarrow 2 \ell 2 \nu)$
for a heavy Higgs boson with Standard Model  $f_{VBF}$ versus the Higgs mass $M_H$ obtained by combining the Run 2 data sets.
The data have been adapted from Fig.~4, bottom right panel, in Ref.~\cite{CMS:2022}. Full black circles correspond to the observed
signal, the red dashed line is the expected Standard Model background together with the 68 \% CL limits (red dotted lines).
The full green  line corresponds to the product of the cross section and branching ratio for our heavy Higgs boson with central mass
$M_H \simeq 730$ GeV. }
\end{figure}
The results are presented as  upper limits on the product of the cross section with  the relevant branching ratio on the production of a
 high mass  scalar resonance. The 68 \% and 95 \% confidence level upper limits 
 on   $\sigma(p  p   \rightarrow  H \rightarrow WW \rightarrow 2 \ell 2 \nu)$ 
are displayed in  Fig.~4 of Ref.~\cite{CMS:2022} for four different scenarios, $f_{VBF} = 0$, $f_{VBF} = 1$, floating $f_{VBF}$ and
Standard Model $f_{VBF} $. Interestingly enough, the CMS Collaboration reported a small excess of data over
the Standard Model background expectations for heavy Higgs boson masses ranging in the interval  500 GeV -   1000 GeV.
Moreover, the signal  hypothesis with the highest local significance  was found in the VBF production mechanism ( $f_{VBF} = 1$)
around the Higgs mass $M_H \simeq 650$ GeV. \\
In order to compare with our theoretical expectations, for definiteness,
 in Fig.~\ref{Fig2} we report  $\sigma(p  p   \rightarrow  H \rightarrow WW \rightarrow 2 \ell 2 \nu)$ as a function of the Higgs mass $M_H$ 
in the case of a heavy scalar resonance with Standard Model-like couplings (Standard Model $f_{VBF}$). The data have been extracted
from  Fig.~4,  bottom right panel, of Ref.~\cite{CMS:2022}. 
Looking at  Fig.~\ref{Fig2} we see that  the observed signals display a sizeable broad excess  with respect to expected Standard Model signal in 
the mass range 600 GeV - 800 GeV.  Clearly, these excesses cannot be accounted for by a heavy scalar resonance with a narrow width.
In addition, for a heavy Standard Model Higgs boson the main production mechanism would be  by gluon-fusion processes for 
$M_H \lesssim  1000$ GeV, so that the resulting production cross section  would led to a  signal greater by at least a factor of two with respect
to the observed signal (see red line in Fig.~4, bottom right panel, of Ref.~\cite{CMS:2022}).
 On the other hand, in our theoretical proposal the heavy Standard Model Higgs boson has a rather large width. So that, the expected main
 signal extends on the mass range 600 GeV - 800 GeV with a broad peak around $M_H \simeq 700$ GeV. Moreover, as we said  in
 Sect.~\ref{s-2}, the main production mechanism is by vector-boson fusion since the gluon-gluon fusion processes are strongly suppressed
 (see Fig.~\ref{Fig1}). To be quantitative, using Eqs.~(\ref{2.4}) and (\ref{2.10}) we may easily evaluate the inclusive production cross
 section for our heavy Higgs boson. The result, displayed in Fig.~\ref{Fig2}, seems to compare reasonable well  to the observed signal
 in the relevant Higgs mass range 600 GeV $  \lesssim M_H  \lesssim$  800 GeV.
It should be emphasised that the  rejection of the  background-only hypothesis in a statistical sense  will depend 
 in general on  the plausibility of the new signal  hypothesis and the degree to which it can describe the data. 
In this respect, the presence of a rather broad excess around  $M_H \simeq 700$ GeV is perfectly consistent
with the fact that our Standard Model heavy Higgs boson has a central mass at   $M_H \simeq 730$ GeV and
a huge width. Moreover, we have estimate that the cumulative effects of the excesses in the mass range 
600 GeV - 800 GeV reach a statistical significance of about eight standard deviations. However, when searching for a new resonance 
somewhere in a possible mass range, the significance of observing a local excess of events must take into account 
the probability of observing such an excess anywhere in the range.  This is the so called "look elsewhere effect"~\cite{Gross:2010}.
Even taking into account the look elsewhere effect, the cumulative statistical significance is at level of five standard deviations.
To obtain more precise statements it should be necessary to implement our heavy Higgs boson in the Monte Carlo numerical
simulations. However, we are aware that the implementation of a heavy resonance with a large width is still problematic.
\section{Conclusions}
\label{s-4}
In our previous  papers we pictured  the Higgs condensate of the Standard Model as a quantum liquid analogous to
the superfluid He II. We found that   the low-lying Higgs condensate excitations behave as two Standard Model Higgs bosons.
The light Higgs boson, identified with the LHC narrow resonance at 125 GeV, turned out to practically indistinguishable from 
the perturbative Standard Model Higgs boson.  As concern the heavy Higgs boson, we  found some evidence in 
our previous phenomenological analysis in the golden channel of the preliminary LHC Run 2 data from ATLAS and CMS Collaborations. 
However, that evidence was not corroborated by the full Run 2 data sets recently released by the ATLAS Collaboration.
Moreover,  considering that the main decay mode of a heavy Higgs boson is the decay into two W vector bosons, the absence 
 of experimental evidences of a heavy Higgs boson in this decay channel constituted a serious problem for our proposal.
In the present note we compared our theoretical proposal to the CMS full Run 2 data set dealing with the search for
 a high mass Higgs boson decaying into a pair of W bosons  in the dileptonic channel.  The main results of the present paper, 
 showed in Fig.~\ref{Fig2}, indicated that our prevision of an additional heavy Standard Model Higgs boson compared 
 in a satisfying way to  the experimental findings.
The agreement between our estimate of the inclusive production cross section and the observed signal in the Higgs mass range
600 GeV $  \lesssim M_H  \lesssim$  800 GeV seems to us particularly significative. Indeed, once one fixes the Higgs masses,
$M_h \simeq 125$ GeV and $M_H \simeq 730$ GeV,  in our theory there are no more free parameters.  Obviously,  to further validate
 our theoretical proposal we must wait for further  LHC Run 2 data, in particular the release by the CMS Collaboration of the full Run 2 dataset
 in the so-called golden channel.

\end{document}